\documentclass[twocolumn,showpacs,showkeys,preprintnumbers,amsmath,amssymb]{revtex4}
\usepackage{graphicx}% Include figure files
\usepackage{dcolumn}% Align table columns on decimal point
\usepackage{bm}% bold math

\begin{document}

%\preprint{PRL Preprint}

\title{Non-quantum liquefaction of coherent gases}
\author{David Novoa, Humberto Michinel, Daniele Tommasini and Mar\'{\i}a I. Rodas-Verde}
\affiliation{Departamento de F\'{\i}sica Aplicada, Facultade de Ciencias de Ourense,\\ 
Universidade de Vigo, As Lagoas s/n, Ourense, ES-32004 Spain.}

\begin{abstract}
%----------------------------   ABSTRACT  ------------------------------------
We show that a gas-to-liquid phase transition at zero temperature may occur 
in a coherent gas of bosons in the presence of competing nonlinear effects.
This situation can take place both in atomic systems like Bose-Einstein Condensates
in alkalii gases with two and three-body interactions of opposite signs, as well 
as in laser beams  which propagate through optical media with Kerr 
(focusing) and higher order (defocusing) nonlinear responses. The liquefaction 
process takes place in absence of any quantum effect and can be formulated in the frame
of a mean field theory, in terms of the minimization of a thermodynamic potential. 
We also show numerically that the effect of linear gain and three-body recombination also provides a 
rich dynamics with the emergence of self-organization behaviour.

\end{abstract}



\maketitle
%------------------------------------- INTRODUCTION -------------------------------------------

\section{Introduction}  
Paraxial propagation of linearly polarized laser beams through transparent optical media with
intensity dependent refractive index is mathematically equivalent to the free evolution of the 
wavefunction order parameter used in the mean-field description of a two-dimensional gas of $N$ 
interacting atoms in a Bose-Einstein Condensate (BEC) at temperature $T=0K$ \cite{bec}. Both systems can
be modeled by identical nonlinear Schr\"odinger equations\cite{book}. For photons in the laser beam, the 
$\chi^{(n)}$ component of the nonlinear optical susceptibility plays the same role as n-body interactions 
between atoms in the cloud and the propagation constant can be identified with a chemical potential for 
the light distribution. As all the photons in a coherent wave are equal, the laser beam can be treated 
on equal foot as any system of $N$ identical interacting bosons at zero temperature\cite{miller77}.  

The previous point of view, which takes into account the equivalence between laser 
beams and BECs of ultracold atoms, has led to an interesting suggestion made by 
Chiao\cite{chiao00}, who recently proposed to verify the superfluidity of coherent 
light, in analogy with degenerate quantum atomic gases. More recently, similar 
concepts have been successfully used to analyze condensation phenomena of nonlinear 
waves\cite{connaughton05} and quantum phase transitions of photons in periodic 
lattices\cite{greentree06}.

In Chiao's model the key point of the analysis is {\em ``same equations, same predictions''} 
and therefore photons from a monochromatic laser source are considered as an ideal bosonic 
gas at zero temperature in which continuous phase transitions (CPT) can take 
place due to long range quantum fluctuations around the ground state\cite{sadchev00}.
These critical phenomena are thus called {\em quantum} phase transitions (QPT)\cite{sondhi97} 
to distinguish them from the standard phase changes which are well-known in classical 
thermodynamics. 

As we will show in this work, CPT may occur in any {\em classical}  
system at zero temperature without long range quantum correlations involved, if opposite 
nonlinear interactions are present. In the case of Chiao's ``superfluid light'' the phase 
transition is produced by effect of a defocusing intensity-dependent refractive 
index\cite{chiao2} and thus a waveguide is used to avoid spreading of the beam (in the same
way as magneto-optical traps are employed to hold atomic BECs). However, as superfluidity may 
occur both in gases and liquids, it cannot be considered as a trace of the presence of 
a liquid state\cite{michinel02,michinel06}; it is also required the appearance of surface 
tension effects\cite{paz-alonso04}. Moreover, in Chiao's model, the interactions between 
particles are repulsive and they cannot drive a gas-liquid transition.

Thus, in this work we will follow the same lines of thought to suggest the possibility of 
obtaining a gas-to-liquid phase transition in a classical gas at $T=0K$ described in 
a mean field theory by the so-called {\em cubic-quintic} (CQ) model with competing nonlinearities. 
Several pioneering works have highlighted the interesting properties of this CQ-model\cite{piekara74}. 
Cavitation, superfluidity and coalescence have been investigated\cite{josserand,josserand_coalescence} 
in the context of liquid He, where the model is a simple approach if nonlocal interactions are not
taken into account. Stable optical vortex solitons and the existence of top-flat 
states have been also reported in optical materials with CQ optical susceptibility\cite{quiroga97,michinel01}. The 
surface tension properties that appear in this system\cite{edmundson95} have been 
considered as a trace of a ``liquid state of light''\cite{michinel02}.

 On the other hand, recent experiments about filamentation of high-power laser pulses in $CS_2$ have shown 
that the CQ nonlinearity is achievable in this material\cite{centurion05}.
It has been also suggested that atomic coherence may be used to induce a giant CQ-like refractive 
index of a Rb gas\cite{michinel06}. Thus, the practical realization of the first ``liquid of light'' state 
as an example of non-quantum liquefaction at zero temperature is close. In BEC systems, a CQ-model can 
be used in the mean field description of an ultracold gas at zero temperature in the presence of Efimov 
states with tunable two- and three-body interactions, which have been recently proposed\cite{zoller}.

%--------------ANALYSIS--------------------%
\section{Mathematical analysis}
The cubic-quintic model describes a coherent bosonic system of $N$ particles
with two- and three-body interactions. The mathematical formulation of the mean field
theory yields a generalised non-linear Schr\"odinger (NLS, also called Gross-Pitaevskii) equation
of the form:

\begin{equation}
\label{NLSE}
i\frac{\partial \Psi }{\partial \eta}+\frac{1}{2}\nabla _{\perp }^{2}\Psi 
+\gamma|\Psi |^{2}\Psi-\delta|\Psi |^{4}\Psi =0,
\end{equation}
If the system modeled by the previous equation is a photon gas, the above NLS describes
paraxial propagation of a continuous linearly-polarized laser beam of wavelength $\lambda$ 
in a  nonlinear medium with a refractive index depending on the intensity $I$ in the 
form $n=n_0+n_2I-n_4I^2$ and the adimensional variables are: $\eta$ the propagation distance 
multiplied by $2\pi/\lambda$, $|\Psi|^2$ the beam irradiance multiplied by the Kerr
coefficient $n_2=\gamma$, $n_4=\delta$ an adimensional parameter indicating the 
strength of the quintic nonlinear optical susceptibility, and 
\( \nabla ^{2}_{\perp }=\partial ^{2}/\partial x^{2}+\partial ^{2}/\partial y^{2} \) 
the transverse Laplacian operator, where $x$ and $y$ are the transverse spatial 
dimensions multiplied by $2\pi\sqrt{2n_0}/\lambda$. 

In the case of a two-dimensional atomic BEC tightly trapped along 
one axis by a parabolic potential of frequency $\nu_\perp$ and thickness 
$r_\perp=\sqrt{\hbar/m\nu_\perp}$, the adimensional variables correspond to: 
$\eta$ the time in units of $\nu_\perp^{-1}$, $|\Psi|^2$ the atomic density multiplied 
by two-body coefficient $\gamma$, $\delta$ an adimensional parameter 
indicating the strength of the three-body interactions, and 
\( \nabla ^{2}_{\perp }=\partial ^{2}/\partial x^{2}+\partial ^{2}/\partial y^{2} \) 
the transverse Laplacian operator, where $x$ and $y$ are the transverse spatial dimensions
divided by $r_\perp$. 

Recent experiments for beam propagation in $CS_2$\cite{centurion05} at $\lambda=800nm$ 
yield the following values for the above parameters in the case of laser beams: 
$n_0=1.6$, $n_2=3\cdot 10^{-15}cm^2/W$ and $n_4=2\cdot 10^{-27}cm^4/W^2$. Other materials 
like air\cite{filamentation} or chalcogenide glasses\cite{smektala00} seem to display 
the C-Q behavior usually accompanied by ionization and non-linear losses. It has been also pointed the 
possibility of engineering this type of optical response by quantum techniques 
which allow to access this nonlinear regime with miliwatt ultrastabilized 
lasers\cite{michinel06}. For atomic BEC systems it has been proposed that a combination 
of two-body (attractive) and three-body (repulsive) elastic interactions can yield 
liquid {\em boselets}\cite{bulgac}. This behaviour can be explained in terms of the
Efimov states\cite{zoller}. However, three-body scattering in BECs has inelastic 
contributions and yields highly nonlinear losses. This means that in the most
general case the coefficient $\delta$ in equation \eqref{NLSE}  may be complex both for laser systems
as well as for atomic gases.

%********************** fig 1  ******************************
\begin{figure}[htbp]
{\centering \resizebox*{1\columnwidth}{!}{\includegraphics{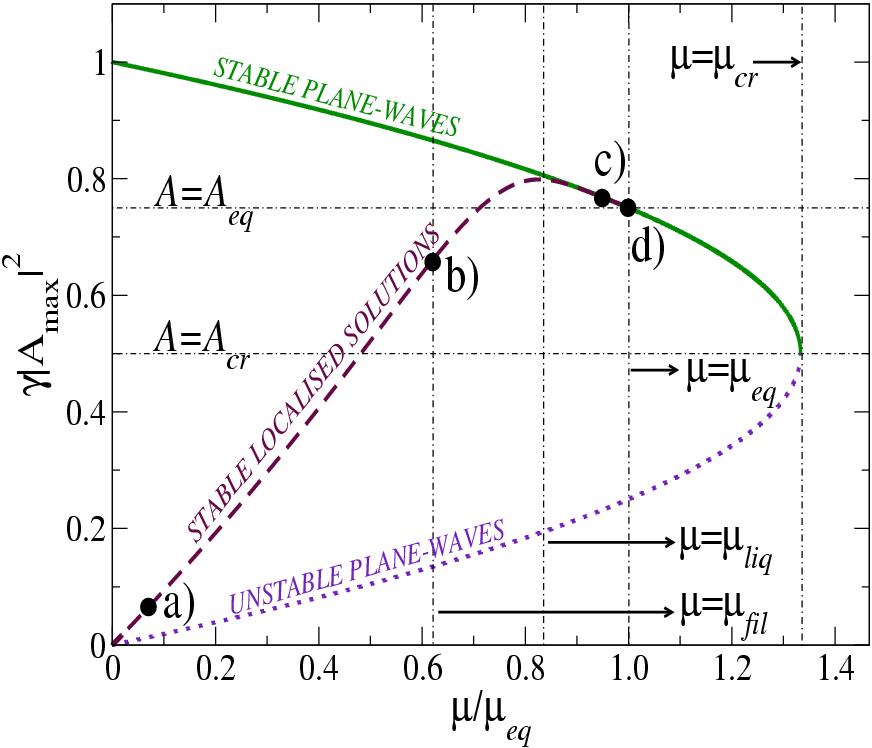}} \par}
\caption {[Color online]
Maximum squared amplitude of different types of stationary solutions of Eq. \eqref{NLSE}.
The continuous and dotted curves correspond to plane waves. The dashed line represents
numerically calculated localized eigenstates. The continuous and dashed curves join at 
$\gamma|A_{max}|^2=0.75$, $|\mu|=|\mu_{eq}|$. Horizontal lines indicates the critical 
values $\gamma|A_{max}|^2=0.5$ and $\gamma|A_{max}|^2=0.75$. Labeled points correspond 
to solitons that are considered as special examples in the text.}
\label{fig1}
\end{figure}
%***************************************************************************

\begin{figure}[htbp]
{\centering \resizebox*{1\columnwidth}{!}{\includegraphics{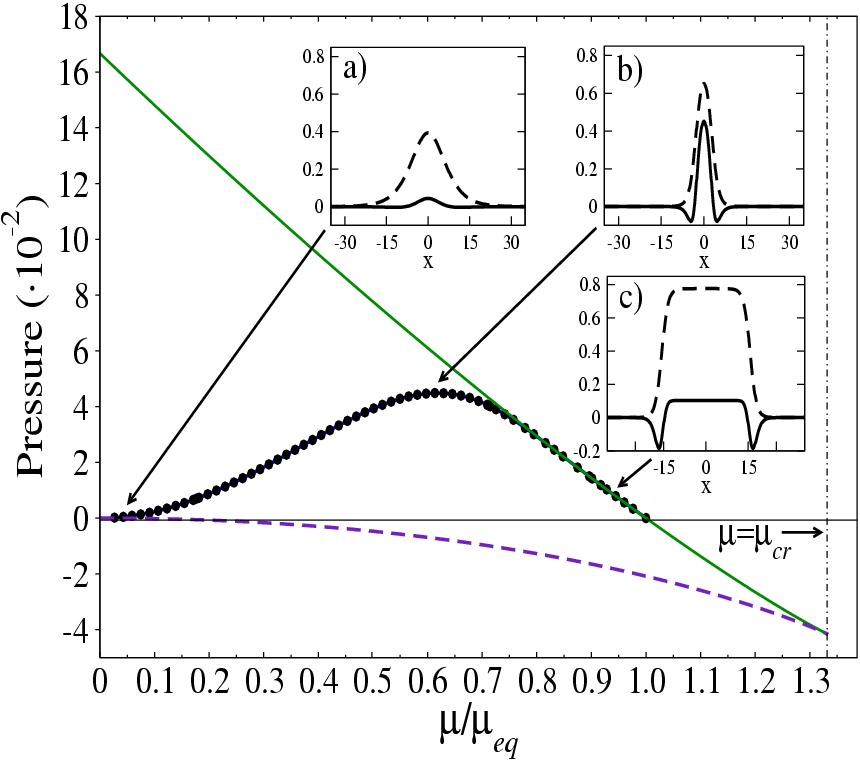}} \par}
\caption{[Color online]
Pressure plots of the plane wave solution branches of Fig.\ref{fig1} (green solid,stable; purple dashed, unstable) 
and pressure at the center ($x=0$) of the eigenstates (thick dotted line) corresponding to different points in the 
dashed line of Fig \ref{fig1}. Insets: Profiles of the eigenstates (dashed) and pressure distribution (solid). 
The x-axis indicates the transverse size of the numerically calculated pressure distributions of different eigenstates
of eq.\eqref{NLSE}. We have multiplied both $\gamma|\Psi|^2$ and $p$ by $10$ and $100$ 
respectively in a) and $p$ by 10 elsewhere.} 
\label{fig2}
\end{figure}

The above CQ-NLSE admits soliton-like solutions of finite size\cite{piekara74} of 
the form $\Psi_A(x,y,\eta)=A(x,y)e^{-i\mu\eta}$, being $\mu$ the propagation constant in the case of
light and the chemical potential for atomic BECs. These solitons can only be calculated numerically 
and coexist with plane waves solutions of constant amplitude $\Psi_A(x,y,\eta)=A e^{-i\mu\eta}$, 
which lead by substitution in Eq.\eqref{NLSE} to $\mu=\delta\vert A\vert^4-\gamma\vert A\vert^2$. In Fig.\ref{fig1} 
we have plotted the maximum value of $\gamma|\Psi|^2$ vs. $\mu$ in units of $\mu_{eq}=-0.1875\gamma^2/\delta$,
for different kinds of stationary solutions of Eq. \eqref{NLSE}. The continuous and dotted 
lines correspond respectively to stable and modulationally unstable plane waves\cite{bespalov66}, whereas the dashed line 
stands for numerically calculated localized eigenstates\cite{michinel04}. It is known\cite{quiroga97} 
that the shape of the solitons (see dashed profiles in Fig.\ref{fig2}) vary from quasi-gaussian shapes for 
low powers to almost square profiles for beam amplitudes close to a certain critical value 
of $\mu=\mu_{eq}$\cite{quiroga97,michinel04}. At this point the size of the solutions tends 
to infinity whereas the amplitude of the beam stabilizes at $|A_{max}|^2=|A_{eq}|^2=0.75\gamma/\delta$. 
We will now clarify the reasons of this behavior by using a thermodynamic model.
%********************** fig 2  ******************************
%***************************************************************************

\section{ Thermodynamic model} 
The appropriate tool to study the equilibrium condition 
as a function of the number of particles $N=\int|\Psi|^2dxdy$ is Landau's grand 
potential
\begin{eqnarray}
\label{Omega}
\Omega&=&H-\mu N=\\ \nonumber
&=&\int dxdy\left[\vert\nabla _{\perp }\Psi\vert^{2}
-\frac{\gamma}{2}|\Psi |^{4}+\frac{\delta}{3} |\Psi |^{6} -\mu |\Psi |^{2}\right],
\end{eqnarray}
where the Lagrange multiplier $\mu$ is the chemical potential for a BEC or
the propagation constant in the case of an optical system. For our two-dimensional 
model the area is $\cal S$ and we have that $p=-\partial\Omega/\partial{\cal S}$ 
is the pressure. The equilibrium configurations are obtained when $\Omega$ is minimized. 
In particular $\partial\Omega/\partial {\cal S}=0$ implies that the pressure 
has to be zero, which for the plane waves of Fig.\ref{fig1} yields
$\partial\Omega/\partial {\cal S}=-\frac{\gamma}{2} |A|^{4}+\frac{\delta}{3} |A|^{6} -\mu |A|^{2}=0$. There
are two possibilities: the trivial case $|A_{eq}|=0$ and a uniform phase,
with $\mu_{eq}=-\frac{3\gamma^2}{16\delta}$ and $|A_{eq}|^2=0.75\gamma/\delta$. 
In the language of field theory, these solutions are the two possible ''vacuum'' 
states of the system. As it can be seen in Fig. \ref{fig1} the two vacua can be ``connected'' by the soliton solutions of the 
dashed line in Fig. \ref{fig1}, which constitute the {\em instantons} of the theory\cite{instantons}.
It is interesting to notice that the non-zero vacuum solution implies a spontaneously 
symmetry breaking of the global phase symmetry, which is preserved in the case with $|A|=0$.

This result about the pressure has already been discussed by authors in ref \cite{josserand} within the framework of 
the Madelung Transformation (MT), which allows one to establish a formal analogy between nonlinear 
optics and classical fluid dynamics\cite{madel_trans}. With the aid of the MT, an analytical expression for the effective 
pressure has been derived\cite{josserand}. Very remarkably, this effective pressure, which is an approximation since
it does not take into account the {\em{quantum-mechanical pressure term}}\cite{madel_trans}, calculated for the plane-wave 
solutions of the CQ-NLSE has the same mathematical expression as the one shown before. Unlike the
 previous works related in the references, we have calculated the effective pressure by means of the potential given by
 eq.\eqref{Omega} which contains all the relevant information about the nonlinear system which is currently being studied. 

While the analysis of the infinite plane-wave solutions has already been considered in the literature \cite{josserand}, 
our discussion of the pressure also applies to the localised soliton solutions, being a central contribution of 
the present work. We also find a significant difference between the low-power quasi-gaussian solutions and the high-power
top-flat solutions of Eq.\eqref{NLSE}. The expression for plane-waves and top-flat eigenstates can be approximated by the 
same ``reduced'' expression for $p$ given by both the MT and the omega density calculation. In fact, we 
have shown in Fig.\ref{fig1} that the two branches of solutions, i.e. the localised solitons and the
stable plane waves, merge as $\mu$ approaches $\mu_{eq}$, having a radius increasing to infinity, thus in this case
it is justified to neglect $\vert\nabla _{\perp }\Psi\vert^{2}$ in Eq.\eqref{Omega}. On the other hand, for the 
quasi-gaussian eigenstates the previous gradient term is not negligible. In this case, we should calculate $p$
numerically in the lack of analytical solutions.

The analysis of the pressure can be refined calculating numerically its distribution 
for different eigenstates as it is plotted in Fig.\ref{fig2}. The pressure of the plane wave branches in Fig.\ref{fig1} is
plotted as a function of $\mu/\mu_{eq}$. Very remarkable is the fact that the pressure of the stable branch is higher than its 
counterpart of the unstable branch. This implies that the stable plane waves free energy density is smaller than the free energy of the
modulationally unstable solutions branch. In Fig.\ref{fig2} it is also plotted the curve of the eigenstates central pressure. 
As it can be seen in the graph, the curve corresponding to the eigenstates is bounded by the two curves of the plane-waves
branches, so that the existence domain of the filaments phase is limited by them. 

In the insets of Fig.\ref{fig2}, we show the shape profiles of the eigenstates (dashed line) superposed with their effective pressure profiles (solid lines) which have been conveniently rescaled to fit in the graph. As it can be appreciated 
in inset a), solitons with $|\mu|\ll|\mu_{eq}|$ have smooth pressure distributions with a central maximum
located at the centroid of the soliton and two negative-valued minima. As the value of $|\mu|$ increases, the soliton profiles and 
their corresponding shapes of $p$ narrow, reaching a minimum width at $\mu/\mu_{eq}=0.5$, which corresponds to a filament 
soliton solution with the same peak amplitude as the ``critical'' plane wave with $|A|=|A_{cr}|$. This plane-wave marks the border 
between stable and modulationally unstable plane waves\cite{josserand}. When the chemical potential reaches the value 
$|\mu|=|\mu_{fil}|$ (see inset b), the absolute maximum of $p$ for eigenstates is obtained, i.e, this filament has the minimum of 
$d\Omega/d\mathcal{S}$ at its centroid. We also consider that close to $|\mu|=|\mu_{liq}|$, the liquid top-flat eigenstates 
begin to exist\cite{michinel02}. In fact, the curve of stable localised solutions in Fig.\ref{fig1} has a maximum in that 
region and both localised solution and stable plane-waves branches seem to merge there. For $|\mu_{liq}|<|\mu|<|\mu_{eq}|$, 
the pressure maximum is located in a flat region (see inset c), tending to zero as $|\mu|$ approaches the critical value $|\mu_{eq}|$, 
point d) in Fig.\ref{fig1}, over which no localized solutions exist\cite{quiroga97}.

%%%%%%%%%%%%%%%%%%%%%%%%%%%%%%%%%%%%%%%%%%%%%%%%%%%%%%%%%%%%%%%%%%%%%%%%%%%%%%%%%%%%%%%%%%%%%%%%%%%%%%%%%%%%%%%%%
\section{Condensation in the presence of linear gain and nonlinear losses} 
In this section, we will provide a set of numerical simulations showing the condensation 
process, i.e., the phase transition from the gaseous phase to an homogeneus coherent 
``liquid'' plane wave solution corresponding to the upper branch in Fig.\ref{fig1}. In order to achieve 
this result, we will perturb an unstable plane wave, corresponding to the lower branch in Fig
\ref{fig1}, with a randomly varying noise. This will produce a filamented phase, made of 
coherent structures whose shape remains qualitatively unchanged, up to smaller scale fluctuations\cite{jordan}.

In order to achieve the non-quantum liquefaction we have included both a
linear incoherent pumping mechanism and nonlinear three-body losses\cite{threebody}. In other words, we have considered 
$\delta=\delta^R+i\delta^I$ and we have introduced a linear gain term $i\Gamma\Psi$ in eq.\eqref{NLSE}. 
This corresponds to a continuous load of particles in the system. Note that in this way, we are describing 
a more realistic non-conservative version of eq.\eqref{NLSE}, which models an experimentally achievable scenario
in the framework of current BEC experiments. In fact, in condensed matter systems it is possible to control the 
load of particles and two and three-body recombination within the coherent atomic cloud\cite{zoller}.
%%********************** fig 3  ******************************
\begin{figure}[htbp]
{\centering
\resizebox*{1\columnwidth}{!}{\includegraphics{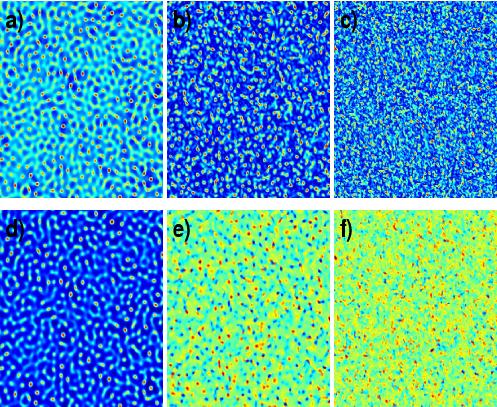}} \par}
\caption {[Color online]
Numerical simulation of the evolution of a set of filaments in presence of linear gain and three-body losses. 
Propagation algorithm parameteres: $\gamma=1$, $\Gamma=10^{-4}$, $\delta=1+0.1i$. Top: pseudocolor maps of 
the amplitude. Bottom: pseudocolor maps of the phase corresponding to the amplitudes above. 
As it can be appreciated in the sequence of snapshots, the initial condition which has a certain 
level of organization, evolves towards a complete disordered situation,i.e., the diluted disordered 
gaseous phase arises. The frames correspond to values of the adimensional 
variable $\eta$: $0$ (a, d), $1000$ (b, e) and $4000$ (c, f).}
\label{fig3}
\end{figure}
%%%%%%%%%%%%%%%%%%%%%%%%%%%%%%%%%%%%%%%%%%%%%%%%%%%%%%%%%%%

On the other hand, in nonlinear optics, this kind of nonlinear models are well-known in the frame of
complex Ginzburg-Landau (G-L) equations used to describe wide-aperture laser cavities\cite{soto}. Although 
it is always possible to control the linear gain introduced in the system, three-body losses are often 
imposed by the nonlinear response of the material, so it is not possible to manage the dissipation terms 
of the system. However, by means of electromagnetic induced transparency techniques, it is possible to 
customize the nonlinear optical response of cold atomic ensembles like Rb\cite{michinel06} 
so that the nonlinear refractive index corresponds to the one given by the modified eq.\eqref{NLSE} analyzed in the 
current section of the paper. Therefore our model, and its predicted phenomenon of liquefaction that we will
demostrate below, can correspond both to realistic BEC and nonlinear optical systems.

In Fig.\ref{fig3}, the initial state consists of an incoherent set of filaments with a randomly varying phase
distribution. Within this apparent disorder, some coherent uncorrelated structures (filaments) exist
and can be observed in Fig.\ref{fig3} a). In this simulation, we have considered a nonlinear parameter 
regime where the linear gain was not enough to compensate the three-body 
dissipative term. As a consequence, certain degree of coherence is lost since 
the filaments disappear and only the noisy background is observed, as shown by snapshot c) in Fig.\ref{fig3}).
%********************** fig 4  ******************************
\begin{figure}[htbp]
{\centering
\resizebox*{1\columnwidth}{!}{\includegraphics{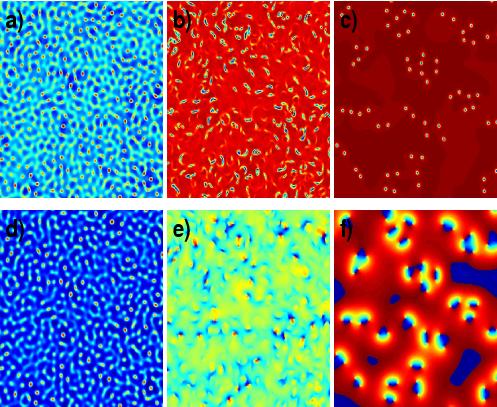}} \par}
\caption {[Color online]
Same as in Fig.\ref{fig3}, but with a modified value of the linear gain $\Gamma=8\cdot 10^{-4}$. Other parameters
as in Fig.\ref{fig3}. In this simulation, the final state is a homogeneous background of coherent liquid with amplitude 
$|A|=|A_{eq}|$, but it is remarkable that the wavefront presents some topological defects (vortices) 
which distort the phase distribution of the resulting liquid plane wave as it can be observed in the phase maps on the 
bottom row. The frames correspond to values of the adimensional 
variable $\eta$: $0$ (a, d), $1000$ (b, e) and $4000$ (c, f).
}
\label{fig4}
\end{figure}
%%%%%%%%%%%%%%%%%%%%%%%%%%%%%%%%%%%%%%%%%%%%%%%%%%%%%%%%%%%%%%
Starting from the same initial condition but increasing the linear gain term over a certain threshold, we have performed the
simulation shown in Fig.\ref{fig4}. In this case, we see that the system evolves towards an homogeneus plane wave by
the combined effect of adding particles to the initial random state and dissipation due to many-body inelastic processes. This
relaxation process is well-known in the context of the complex G-L equations and is attributted to the non-conservative nature of
the model\cite{Ginz_Land}. Nevertheless, as our theory predicts, the system will tend to form a particular plane wave 
of the stable ``liquid'' upper branch of Fig.\ref{fig1}, the one with zero pressure. This illustrates the tendency of the
system to reach the non-zero vacuum state with $|A|^2=3\gamma/4\delta$.
%********************** fig 5  ******************************
\begin{figure}[htbp]
{\centering \resizebox*{1\columnwidth}{!}{\includegraphics{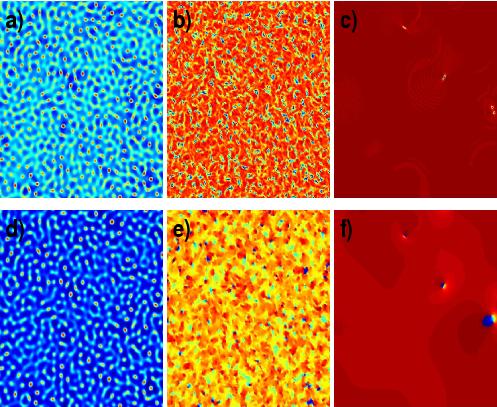}}
\par}
\caption {[Color online]
Numerical simulation of the evolution of a set of filaments in presence
of C-Q nonlinear gain and loss. The linear gain term used in the previous simulations has been replaced
here by a nonlinear gain term $i\chi|\Psi|^2$, where $\chi=0.1$. Other parameters are: $\gamma=1$, $\delta=1+0.1i$. 
Top: pseudocolor maps of the amplitude. Bottom: pseucolor maps of the phase corresponding to the amplitudes above. 
Qualitatively, the same results of Fig.\ref{fig4} are obtained. However, it is important to note that in 
this case, the robust topological structures with nonzero vorticity which appear during the dynamical 
liquefaction process are annihilated, resulting in a final homogeneous background with amplitude 
$|A|=|A_{eq}|$ and constant phase distribution, as it can be seen in snapshot f). The frames correspond 
to values of the adimensional variable $\eta$: $0$ (a, d), $1000$ (b, e) and $4000$ (c, f).
}
\label{fig5}
\end{figure}
%%%%%%%%%%%%%%%%%%%%%%%%%%%%%%%%%%%%%%%%%%%%%%%%%%%%%%%%%%%%%%%%%
Very remarkably, during the process pairs of vortex-antivortex with topological charges $m_v=1,m_{av}=-1$ are
formed (see bottom phase maps in Fig.\ref{fig4}), so that the constant phase of the emerging coherent wave remains hidden by the
overlapping of the different vortex rotating phase-distributions. Vortices are very robust topological structures\cite{paz-alonso04}
and in our simulations they remain stable as far as we could follow the numerical simulations.

Finally, we have considered the effect of replacing the linear gain term by a nonlinear gain term of the form $i\chi|\Psi|^2$. In this
situation, the numerical results are qualitatively the same as in Fig.\ref{fig4}, although it can be observed that in the
last stage of the evolution the vortices are annihilated. In snapshot f) of Fig.\ref{fig4}, it can be seen how the underlying plane
wave, which emerges after the dynamical process as described above, has a homogeneus phase distribution.
We think that this result is a very interesting example of a self-organization process, where coherence is produced from disorder, 
with evident practical applications.

%----------------------- CONCLUSIONS ----------------------------------------
\section{Conclusions}
We have shown in the present work that a system of $N$ equal bosons in a system with competing 
nonlinearities can undergo a phase transition from a gas to a liquid state. The process takes 
place at zero temperature without any quantum effect and it is only ruled by nonlinear interactions. 
Finally, we have shown that a cubic-quintic medium with complex susceptibilities exhibiting linear gain 
and nonlinear losses, will tend to produce a homogeneous phase liquid distribution starting from a 
collection of non-correlated filaments. This opens the door for experiments in the field of 
BEC systems in ultracold gases.

\section{Acknowledgments} H. M. thanks A. J. Legget and A. Ferrando for useful 
discussions and Y. Castin, J. Ho and G. V. Shlyapnikov for warm hospitality 
at Inst. Henri Poincar\'e. This work was supported by MEC, Spain 
(projects FIS2006-04190 and FIS2007-62560) Xunta de Galicia (project PGIDIT04TIC383001PR
and D.N. grant from Conseller\'{i}a de Educaci\'on, Xunta de Galicia).

%----------------------------------------------------------------------------


\begin{thebibliography}{99}

\bibitem{bec}{A.J. Leggett, ``{\em Quantum Liquids: Bose Condensation and Cooper Pairing 
in Condensed-Matter Systems}'', (Oxford University Press, N.Y. 2006). }

\bibitem{book}{C. Sulem and P.L. Sulem, ``{\em Nonlinear Schr\"odinger Equations: 
Self-Focusing and Wave Collapse}'', (Springer-Verlag, N.Y. 1999).}

\bibitem{miller77}{M. D. Miller, L. H. Nosanow, and L. J. Parish, 
\prb {\bf 15}, 214-229 (1977).}

\bibitem{chiao00} R. Y. Chiao,
%''Bogoliubov dispersion relation for a 'photon fluid': is this a superfluid?
\oc {\bf 179}, 157--166 (2000).

\bibitem{connaughton05} C. Connaughton {\em et al.},
%''Condensation of classical nonlinear waves''
\prl {\bf 95}, 263901 (2005). 

\bibitem{greentree06}{ A. D. Greentree, et. al., Nature Phys., {\bf 2}, 856-861 (2006).}

\bibitem{sadchev00}{S. Sadchev, Science {\bf 288}, 475-479 (2000).}

\bibitem{sondhi97}{S. L. Sondhi, {\em et al.}, \rmp {\bf 69}, 315-333 (1997).}

\bibitem{chiao2} E. L. Bolda, R. Y. Chiao and W. H. Zurek, 
%''Dissipative Optical Flow in a Nonlinear Fabry-Perot cavity''
\prl {\bf 86}, 416--419 (2001); 
R. Y. Chiao {\em et al.},
%''Two-dimensional 'photon fluid': effective photon-photon interaction and physical realizations.
J. Phys. B: At. Mol. Opt. Phys {\bf 37}, S81--S89 (2004); 
R. Y. Chiao {\em et al.},
%''Effective Photon-Photon interaction in a two-dimensional ``photon fluid''
\pra {\bf 69}, 063816 (2004); 
A. Tanzini and S.P. Sorella
%''Bose-Einstein condensation and superfluidity of a weakly-interacting photon gas in a nonlinear Fabry-Perot cavity''
\pl A {\bf 263}, 43--47 (1999). 
	
\bibitem{michinel02} H. Michinel {\em et al.},
%''Liquid light condensates''
\pre {\bf 65}, 066604 (2002).

\bibitem{michinel06} H. Michinel, M. J. Paz-Alonso and V. M. P\'erez Garc\'{i}a,
%''Turning light into a liquid via atomic coherence''
\prl {\bf 96}, 023903 (2006).

\bibitem{paz-alonso04} M. J. Paz-Alonso {\em et al.},
%''Collisional dynamics of vortices in light condensates''
\pre {\bf 69 }, 056601 (2004).

\bibitem{piekara74} A. H. Piekara, J. S. Moore, and M. S. Feld, 
%``Analysis of self-trapping using the wave equation with high-order nonlinear electric permitivity,'' 
\pra {\bf 9}, 1403--1407 (1974).

\bibitem{josserand} C. Josserand, Y. Pomeau and S. Rica,
%''Cavitation versus vortex nucleation in a Superfluid model''
\prl {\bf 75}, 3150--3153 (1995);
C. Josserand,
%''Cavitation induced by explosion in an ideal fluid model''
\pre {\bf 60}, 482--491 (1999);

\bibitem{josserand_coalescence} C. Josserand and Sergio Rica,
%''Coalescence and droplets in the Subcritical Nonlinear Schrodinger equation''
\prl {\bf 78}, 1215--1218 (1997).

\bibitem{quiroga97} M. Quiroga-Teixeiro and H. Michinel, 
%``Stable azimutal stationary state in quintic nonlinear optical media,'' 
\josab {\bf 14}, 2004--2009 (1997); D. Mihalache, {\em et al.}, \prl {\bf 88}, 073902 (2002).

\bibitem{michinel01} H. Michinel, J. Campo-T\'aboas, M. L. Quiroga-Teixeiro, J. R. Salgueiro, and R. Garc\'{\'i}a Fern\'andez,
J. Opt. B: Quantum Semiclass. Opt. {\bf 3}, 314 (2001); M. J. Paz-Alonso and H. Michinel, Phys. Rev. Lett. {\bf 94}, 093901 (2005).


\bibitem{edmundson95} D. E. Edmundson and R. H. Enns, 
%``Particlelike nature of colliding three-dimensional optical solitons,'' 
\pra {\bf 51}, 2491--2498 (1995);

\bibitem{centurion05} M. Centurion {\em et al.},
%''Dynamics of filament formation in a Kerr medium''
\pra {\bf 71}, 063811 (2005).

\bibitem{zoller} H. P. Buchler, A. Micheli, and P. Zoller P,
Nature Phys. {\bf 3}, 726-731 (2007).

\bibitem{filamentation}S. Tzortzakis {\em et al.},
%''Breakup and fusion of Self-guided Femtosecond light pulses in air''
\prl {\bf 86}, 24 (2001); 
A. Couairon,
%''Dynamics of femtosecond filamentation from saturation of self-focusing lases pulses''
\pra {\bf 68}, 015801 (2003);
S. Skupin {\em et al.},
%''Filamentation of femtosecond light pulses in the air: Turbulent cells versus long-range clusters''
\pre {\bf 70}, 046602 (2004);
L. Berg\'e {\em et al.},
%''Multiple Filamentation of Terawatt laser pulses in air''
\prl {\bf 92}, 22 (2004);
C. Ruiz {\em et al.},
%''Observation of Spontaneous Self-Channeling of Light in Air below rhe Collapse Threshold''
\prl {\bf 95}, 053905 (2005).

\bibitem{smektala00} F.Smektala, C.Quemard, V.Couderc, and A.Barthelemy, 
%``Non-linear optical properties of chalcogenide glasses measured by Z-scan,''
J. Non-Cryst. Sol., {\bf 274}, 232-237 (2000).

\bibitem{jordan} R. Jordan and C. Josserand, 
% self-organization in nonlinear wave turbulence
\pre {\bf 61}, 2 (2000).

\bibitem{bulgac} A. Bulgac,
%''Dilute quantum droplets''
\prl {\bf 89}, 050402 (2002); 
A. Gammal {\em et al.},
%''Liquid-gas phase transition in Bose-Einstein condensates with time evolution''
\pra {\bf 61}, 051602(R) (2000); H.-W. Hammer and D.T. Son, \prl {\bf 93}, 250408 (2004).

\bibitem{michinel04}{H. Michinel, J. R. Salgueiro, and M.J. Paz-Alonso, \pre {\bf 70}, 066605 (2004).}

\bibitem{instantons}{T. Schafer and E. V. Shuryak, \rmp {\bf 70}, 323-425 (1998).}

\bibitem{madel_trans}{C. Coste, Eur. Phys. J. B. {\bf 1}, 245-253 (1998).}

\bibitem{mewes97}{M. -O. Mewes, M. R. Andrews, D. M. Kurn, D. S. Durfee, 
C. G. Townsend and W. Ketterle, Phys. Rev. Lett. {\bf 78}, 582 (1997).}

\bibitem{bespalov66} V. I. Bespalov and V. I. Talanov,
%''Filamentary structure of light beams in nonlinear liquids''
JETP Lett. {\bf 3}, 307 (1966).

\bibitem{centurion06} M. Centurion, {\em et al.}, \pra {\bf 74}, 069902(E) (2006).

\bibitem{threebody} E. Braaten and H.W. Hammer,
%''Three body recombination into deep bound states in a bose gas with large scattering length''
\prl {\bf 87}, 16 (2001).

\bibitem{Ginz_Land} LC. Crasovan, BA. Malomed and D. Mihalache, 
%''Stable vortex solitons in the complex Ginzburg-Landau equation''%
\pre {\bf 63}, 016605 (2001);
P. Grelu, J. Soto-Crespo and N. Akhmediev,
%''Light bullets and dynamic pattern formation in nonlinear dissipative systems''
Opt. Express {\bf 13}, 23, pp: 9352-9360 (2005)
S. Mancas, SR. Choudhury,
%''Bifurcations and competing coherent structures in the cubic-quintic Ginzburg-Landau equation: plane wave (CW) solutions.
Chaos, Solitons and Fractals, {\bf 27}, 5, pp: 1256-1271 (2006)

\bibitem{soto}{J.M. Soto-Crespo, P. Grelu, and N. Akhmediev, Opt. Express {\bf 14}, 4013-4025 (2006).}

\end{thebibliography}
\end{document}